\documentclass[twocolumn,prd,amsmath,amssymb]{revtex4-1}
\usepackage{graphicx}
\usepackage{amsmath}
\usepackage{dcolumn}
\usepackage{bm}
\usepackage{epsfig}
\usepackage{amssymb}

\setcitestyle{authoryear,round,semicolon}	
\usepackage{url}
\usepackage[usenames,dvipsnames]{color}
\usepackage{multirow}
\usepackage{nicefrac}
\usepackage{subfigure}

\newcommand{\hess}{H.E.S.S.}
\newcommand{\hessj}{HESS\,J}
\newcommand{\psrj}{PSR\,J}
\newcommand{\vheg}{VHE $\gamma$-ray}

\begin{document}
   \title{Predicting the X-ray flux of evolved pulsar wind nebulae based on \vheg\ observations}

   \author{M. Mayer$^{1\mathrm{,}2}$, J. Brucker$^{1}$, M. Holler$^{1\mathrm{,}2}$, I. Jung$^{1}$, K. Valerius$^{1}$ and C. Stegmann$^{3\mathrm{,}2}$}

   \affiliation{$^{1}$Erlangen Centre for Astroparticle Physics (ECAP), Universit\"at Erlangen-N\"urnberg, Erwin-Rommel-Str. 1, D-91058 Erlangen, Germany \\
   $^2$Institut f\"ur Physik und Astronomie, Universit\"at Potsdam, Karl-Liebknecht-Str. 24/25, D-14476 Potsdam-Golm, Germany \\
   $^3$DESY, Platanenallee 6, D-15738 Zeuthen, Germany }

  \date{\today}

\begin{abstract}
{\it Context.} Energetic pulsars power winds of relativistic leptons which produce photon nebulae (so-called pulsar wind nebulae, PWNe) detectable across the electromagnetic spectrum up to energies of several TeV. The spectral energy distribution has a double-humped structure: the first hump lies in the X-ray regime, the second in the $\gamma$-ray range. The X-ray emission is generally understood as synchrotron radiation by highly energetic leptons, the $\gamma$-ray emission as Inverse Compton scattering of energetic leptons with ambient photon fields. The evolution of the spectral energy distribution is influenced by the time-dependent spin-down of the pulsar and the decrease of the magnetic field strength with time. Thus, the present spectral appearance of a PWN depends on the age of the pulsar: while young PWNe are bright in X-rays and $\gamma$-rays, the X-ray emission of evolved PWNe is suppressed. Hence, evolved pulsar wind nebulae may offer an explanation of the nature of some of the unidentified VHE (very high-energy, $E >100\,\mathrm{GeV}$) $\gamma$-ray sources not yet associated with a counterpart at other wavelengths.\newline
{\it Aims.} The purpose of this work is to develop a model which allows to calculate the expected X-ray fluxes of unidentified \vheg\ sources considered to be PWN candidates. Such an estimate may help to evaluate the prospects of detecting the X-ray signal in deep observations with current X-ray observatories in future studies.\newline
{\it Methods.} We present a time-dependent leptonic model which predicts the broad-band emission of a PWN according to the characteristics of its pulsar. The values of the free parameters of the model are determined by a fit to observational VHE $\gamma$-ray data. For a sample of representative PWNe, the resulting model predictions in the X-ray and $\gamma$-ray range are compared to observations.\newline
{\it Results.} The comparison shows that the energy flux of the X-ray emission of identified PWNe from different states of evolution can be roughly predicted by the model. This implies the possibility of an estimate of the non-thermal X-ray emission of unidentified VHE $\gamma$-ray sources in case of an evolved PWN scenario.
\end{abstract}


   \maketitle

\section{Introduction}
\label{sec:introduction}
Due to a new generation of Imaging Atmospheric Cherenkov Telescopes (IACTs), the number of detected Galactic sources emitting VHE (very high-energy, $E >100\,\mathrm{GeV}$) $\gamma$-rays has increased significantly during the last decade. Pulsar wind nebulae (PWNe) form the most abundant class among these sources. Such nebulae are usually associated with the non-thermal emission from a magnetized plasma of relativistic particles fed by an energetic pulsar. In current models, the plasma is thought to consist mainly of energetic leptons \citep[see, e.g.,][]{Gaensler2006} which emit non-thermal radiation over a wide energy range. Interacting with magnetic fields, the leptons produce synchrotron radiation up to several MeV. In addition, low-energy photons, e.g. from the cosmic microwave background (CMB), can be up-scattered by the energetic leptons to very high energies via the Inverse Compton effect. Therefore, the emission in X-rays and \mbox{VHE $\gamma$-rays} is tightly linked, emerging from the same lepton population \citep[see, e.g.,][]{Gelfand2009}. The second largest population of Galactic \vheg\ sources consists of unidentified sources without an unambiguous counterpart at other wavelengths \citep{Aharonian2008}. However, in many cases an energetic pulsar can be found in the vicinity, suggesting a possible connection between these unidentified objects and pulsar wind nebulae. Provided that the rotational period and its first time derivative can be measured, e.g. by radio observations, the spin-down energy loss of the pulsar can be estimated and hence the viability of the pulsar as an energy source of the nebula can be investigated. In addition, in the PWN scenario of broad-band emission by energetic leptons a spatial association of the \vheg\ source with an \mbox{X-ray} nebula counterpart is expected. There are mainly two issues complicating this identification scheme: In some cases, PWNe are slightly displaced from the pulsar, which may result from an interaction with the supernova remnant reverse shock \citep[see, e.g.,][]{Blondin2001} and from a proper motion of the pulsar, gained from a kick at its birth \citep[e.g.][]{vanderSwaluw2004}. Furthermore, in particular for older systems the \mbox{X-ray} emission becomes fainter and hence harder to detect since the energetic leptons injected during earlier epochs have been cooled and, at the same time, the supply of fresh leptons is reduced. Moreover, the synchrotron emission by the freshly injected leptons is suppressed because the magnetic field strength decreases with time. Since the accumulated less energetic leptons can still produce \vheg s via Inverse Compton (IC) scattering, such evolved PWNe have been proposed \citep{deJager2009} as an explanation of some of the as yet unidentified \vheg\ sources.
\newline
In this work, we introduce a time-dependent leptonic model of the non-thermal emission of PWNe (Section\;2) and apply the model to PWNe of different evolutionary states (Section\;3). For each individual source, the free parameters are fixed by fitting the model to the \mbox{VHE $\gamma$-ray} data. Subsequently, we show that the fitted model allows a rough prediction of the X-ray emission of these objects. Hence, in future studies it may serve as a means to estimate the X-ray flux of unidentified \vheg\ sources in a PWN scenario, allowing to evaluate the prospects of detection in deep observations with current X-ray observatories.

\section{The Model}
\label{sec:model}
\begin{figure}[!h]
\centering
\includegraphics[width=\linewidth]{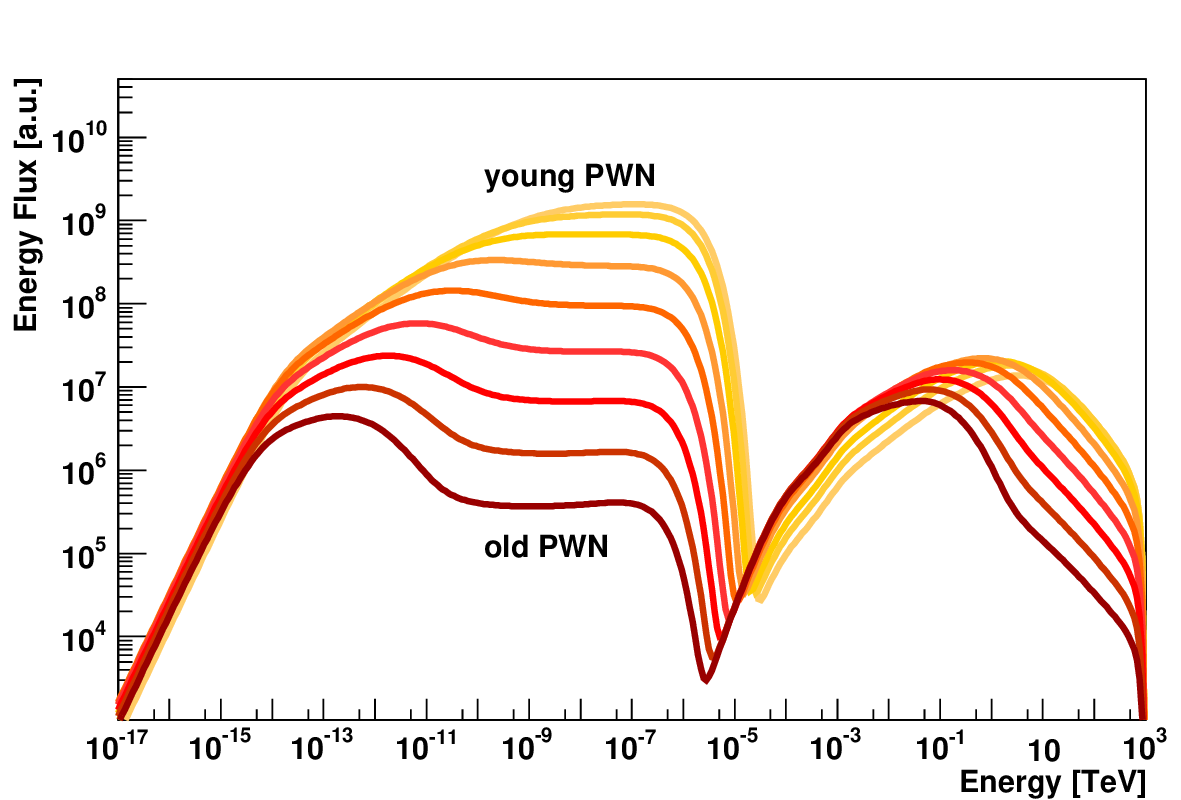}
\caption{Typical evolution of the modeled spectral energy distribution of a PWN with time. The color scale represents the age of the PWN, starting with a young system ($500$\,years, yellow) and proceeding in equidistant steps on a logarithmic time scale to an old system ($150$\,kyr, dark red).}
\label{fig:sed_evo}
\end{figure}

In this Section, we introduce a leptonic model describing the time evolution of the non-thermal radiation from PWNe. The time dependence of the energy output $\dot{E}$ of the pulsar, which derives from the slow-down of the rotation, has to be taken into account:
\begin{equation}
\dot{E} = -\frac{\mathrm{d}E_\mathrm{rot}}{\mathrm{d}t}\mathrm{.}
\end{equation}
Following \citet{Pacini1973}, the energy output evolves with time as
\begin{equation}
\dot{E}(t) = \dot{E}_0 \left(1+\frac{t}{\tau_{0}}\right)^{-\frac{n+1}{n-1}}\mathrm{,}
\end{equation} 
where $\dot{E}_0 = \dot{E}(t=0)$, $\tau_0$ denotes the spin-down timescale of the pulsar and $n$ the braking index. The latter has been measured only for a few young pulsars \cite[see, e.g.,][and references therein]{Magalhaes2012}. Such a measurement exists, for instance, for \mbox{PSR B1509$-$58}, which is one of the sample pulsars discussed in Section\;3. Thus, for the modeling of the PWN associated with \mbox{PSR B1509$-$58} we used the measured value of $n=2.839$ \citep{Livingstone2007}. For the other cases we adopted $n=3$ \citep{ManchesterTaylor1977}, corresponding to spin-down via magnetic dipole radiation. The spin-down timescale $\tau_{0}$ is defined by
\begin{equation}
\tau_{0} = \frac{2\tau_{\mathrm{c}}}{n-1}\left(\frac{P_0}{P}\right)^{n-1}\mathrm{,}
\end{equation}
with $P_0$ and $P$ being the initial and the current period, respectively, and $\tau_\mathrm{c} = P/(2\dot{P})$ the characteristic age of the pulsar ($\dot{P}$ denoting the time derivative of the rotational period).
For $n = 3$ and $P_0\ll P$ the present true age $T$ of the pulsar corresponds to the characteristic age $\tau_\mathrm{c}$, whereas for other cases it can be calculated as
\begin{equation}
T = \frac{P}{(n-1)\dot{P}}\left(1-\left(\frac{P_0}{P}\right)^{n-1}\right)\mathrm{.}
\end{equation}
$P$ and $\dot{P}$ can usually be derived from radio observations, while the initial period $P_0$ will be treated as a free parameter of our model. \newline
In the following, the evolution of the non-thermal emission of a PWN is calculated in discrete time steps with an adaptive step size of $\delta t$. In each time step only a fractional amount $\Delta E_\mathrm{p}$ of the energy output of the pulsar is converted into relativistic leptons, i.e. electrons and positrons. Assuming the corresponding conversion efficiency $\eta$ to be constant over time, $\Delta E_\mathrm{p}(t)$ is determined as
\begin{equation}
\Delta E_\mathrm{p}(t) = \eta \int_t^{t+\delta t} \dot{E}(t')\mathrm{d}t'
\end{equation}
for a time interval $[t,t+\delta t]$ and with $\eta\in [0,1]$. 
The conversion efficiency is strongly correlated with $P_0$, such that the quality of the fit does not benefit from an additional free parameter. Therefore, $\eta$ is fixed to the value of 0.3, which is in agreement with e.g. the modeling results for MSH\,15$-$5\emph{2} carried out by \cite{Schoeck2010} and \cite{Zhang2008}. We transferred this value to the other selected PWNe.\newline
We assume that the differential energy spectrum of the injected leptons can be described by a simple power law 
\begin{equation}
\frac{\mathrm{d}N_\mathrm{inj}}{\mathrm{d}E}(E,t)=\Phi_0(t)\left(\frac{E}{1\,\mathrm{TeV}}\right)^{-2}\mathrm{.}
\end{equation}
Assuming that the spectral shape does not change within a time bin, $\Phi_0(t)$, denoting the normalization of the distribution at $1\,\mathrm{TeV}$, can be calculated by integrating the injection spectrum over energy:
\begin{equation}
\Delta E_\mathrm{p}(t) \stackrel{!}{=} \int_{E_{\mathrm{min}}}^{E_{\mathrm{max}}}\frac{\mathrm{d}N_\mathrm{inj}}{\mathrm{d}E}(E,t)\,\mathrm{d}E\,\mathrm{.}
\end{equation}
We only consider leptons injected into the pulsar wind in the range between $E_{\mathrm{min}}=0.1$\,TeV and $E_{\mathrm{max}}=1000\,\mathrm{TeV}$, well suited to accomodate the \mbox{VHE $\gamma$-rays} as well as the X-ray emission from PWNe. Leptons with energies outside this range do not significantly contribute to the emission in the considered photon wavebands. Given a differential yield of leptons $\mathrm{d}N(E,t-\delta t)/\mathrm{d}E$ with energy $E$ at a time $t-\delta t$ the number of leptons remaining after cooling at the time $t$ can be calculated. Following \citet{Zhang2008}, the cooling of the lepton population during a time step $\delta t$ is implemented in the model by means of an exponential function:
 \begin{equation}
 \frac{\mathrm{d}N_\mathrm{cooled}}{\mathrm{d}E}(E,t)=\frac{\mathrm{d}N}{\mathrm{d}E}(E,t-\delta t)\cdot\exp\left(-\frac{\delta t}{\tau_{\mathrm{eff}}(E,t)}\right)\mathrm{.}
\end{equation}
This approach uses an effective cooling timescale \mbox{$\tau_\mathrm{eff}^{-1} = \tau_\mathrm{syn}^{-1} + \tau_\mathrm{esc}^{-1} + \tau_\mathrm{ad}^{-1}$} taking into account synchrotron, escape and adiabatic energy losses. In general, the cooling timescale of a particle with energy $E$ and a current energy loss rate $\dot{E}_\mathrm{p}$ is defined by $\tau=-\frac{E}{\dot{E}_\mathrm{p}}$. The synchrotron and escape cooling time scales $\tau_{\mathrm{syn}}$ and $\tau_\mathrm{esc}$ are likewise adopted from \citet{Zhang2008}:
\begin{equation}
\tau_{\mathrm{syn}}(E,t) = 12.5\cdot\left[\frac{B(t)}{10\,\mathrm{\mu G}}\right]^{-2}\cdot\left[\frac{E}{10\,\mathrm{TeV}}\right]^{-1}\,\mathrm{kyr}
\end{equation}
\begin{equation}
\tau_{\mathrm{esc}}(E,t) = 34\cdot\left[\frac{B(t)}{10\,\mathrm{\mu G}}\right]\cdot\left[\frac{E}{10\,\mathrm{TeV}}\right]^{-1}\cdot\left[\frac{R(t)}{1\,\mathrm{pc}}\right]^{2}\,\mathrm{kyr,}
\end{equation} 
where $R(t)$ and $B(t)$ describe the time evolution of the PWN radius and the magnetic field strength inside the PWN, respectively. For evolved PWNe, $R(t)$ is given by \citep[see][and references therein]{Gaensler2006}:
\begin{equation}
R(t) = \quad
\begin{cases}
\quad a\cdot t^{\nicefrac{11}{15}} & \textrm{for} \quad t < \tau_0 \\
\quad b\cdot t^{\nicefrac{3}{10}} & \textrm{for} \quad t \ge \tau_0
\end{cases}  \mathrm{,}
\label{eq:radius}
\end{equation}
where the coefficients $a$ and $b$ can be calculated using the present-day size of the PWN. Based on the radius evolution, we can calculate the timescale for adiabatic energy losses following \cite{Harding1992}:
\begin{equation}
\frac{\mathrm{d}E_\mathrm{ad}}{\mathrm{d}t}=-\frac{E}{3}\nabla \vec{v}_\mathrm{\bot}(R)=\dot{E}_\mathrm{p}
\label{eq:adiab}
\end{equation}
with $\vec{v}_\mathrm{\bot}(R)$ being the radial component of the particle velocity. In general, its divergence can be calculated to 
\begin{align}
\nabla\vec{v}_\mathrm{\bot}(R) &=\frac{1}{R^{2}}\cdot\frac{\partial (R^{2}\vec{v}_\mathrm{\bot})}{\partial R}\\
&=\frac{1}{R(t)^{2}}\cdot\frac{\partial (R(t)^{2} \vec{v}_\mathrm{\bot}(t))}{\partial t}\cdot\frac{\partial t}{\partial R}
\end{align}
\begin{equation}
\Rightarrow\tau_{\mathrm{ad}} = -\frac{E}{\dot{E_\mathrm{p}}} = \quad
\begin{cases}
\quad\nicefrac{45}{18}\cdot t & \textrm{for} \quad t < \tau_0 \\
\quad -30\cdot t & \textrm{for} \quad t \ge \tau_0
\end{cases}
\end{equation}
This results in an adiabatic energy gain of particles in older PWNe. However, this energy gain is negligible due to the significantly larger time scale compared to the other processes. The magnetic field strength $B(t)$ is adapted from \cite{Zhang2008}:
\begin{equation}
B(t) =\frac{B_0}{1+\left(\nicefrac{t}{\tau_0}\right)^{\alpha}}+B_\mathrm{ISM}\,\mathrm{.}
\end{equation}
$B_\mathrm{ISM}$ represents a time-independent component of $3\,\mathrm{\mu G}$ to account for the magnetic field strength of the ambient medium. Assuming the conservation of magnetic flux density for large time scales ($t\gg\tau_0$) implies $\alpha=0.6$. Finally, $B_0$, the initial magnetic field strength inside the PWN, is a free parameter. All in all, the model has two free parameters ($P_0$ and $B_0$) defining the starting conditions of the PWN evolution.\newline Having established the framework for cooling and injection processes, we can calculate the number of leptons with energy $E$ present in the nebula at a time $t+\delta t$. This number comprises leptons injected and cooled until time $t$ as well as freshly injected leptons between $t$ and $t+\delta t$:
\begin{equation}
\frac{\mathrm{d}N}{\mathrm{d}E}(E,t+\delta t) = \frac{\mathrm{d}N_\mathrm{cooled}}{\mathrm{d}E}(E,t) +  \frac{\mathrm{d}N_\mathrm{inj}}{\mathrm{d}E}(E,t+\delta t) .\label{eq:amount}
\end{equation}  
By iteratively evaluating Eq.~\ref{eq:amount}, it is possible to determine the energy distribution of the leptons inside the PWN at an arbitrary time. Based on this distribution, the corresponding photon population can be calculated, with synchrotron radiation and IC scattering as the most relevant emission processes in the considered energy range. A detailed account of these mechanisms can be found in \citet{Blumenthal1970}. We neglect a synchrotron self-Compton (SSC) scattering component in the VHE $\gamma$-ray spectrum, since this work is focused on evolved pulsars, whereas SSC is relevant mostly for the highly magnetized PWNe of very young and energetic pulsars, e.g. the Crab Nebula \citep{Meyer2010}. The target photon fields considered for IC scattering  -- CMB, starlight and infrared photons -- are adopted from the {\sc galprop} code \citep{Porter2005}.\newline
As a first step, we can use the model to study the development of the spectral energy distribution (SED) with progressing age for a generic PWN system. The SED shown in Fig.~\ref{fig:sed_evo} is based on the characteristics of the pulsar \psrj 1826$-$1334 and its nebula (see Table \ref{tab:sources}), representing an example of an evolved PWN. The free parameters are exemplarily set to $P_0 = 30$\,ms and $B_0= 50\,\mu$G.
\begin{figure}[!h]
\centering
\includegraphics[width=\linewidth]{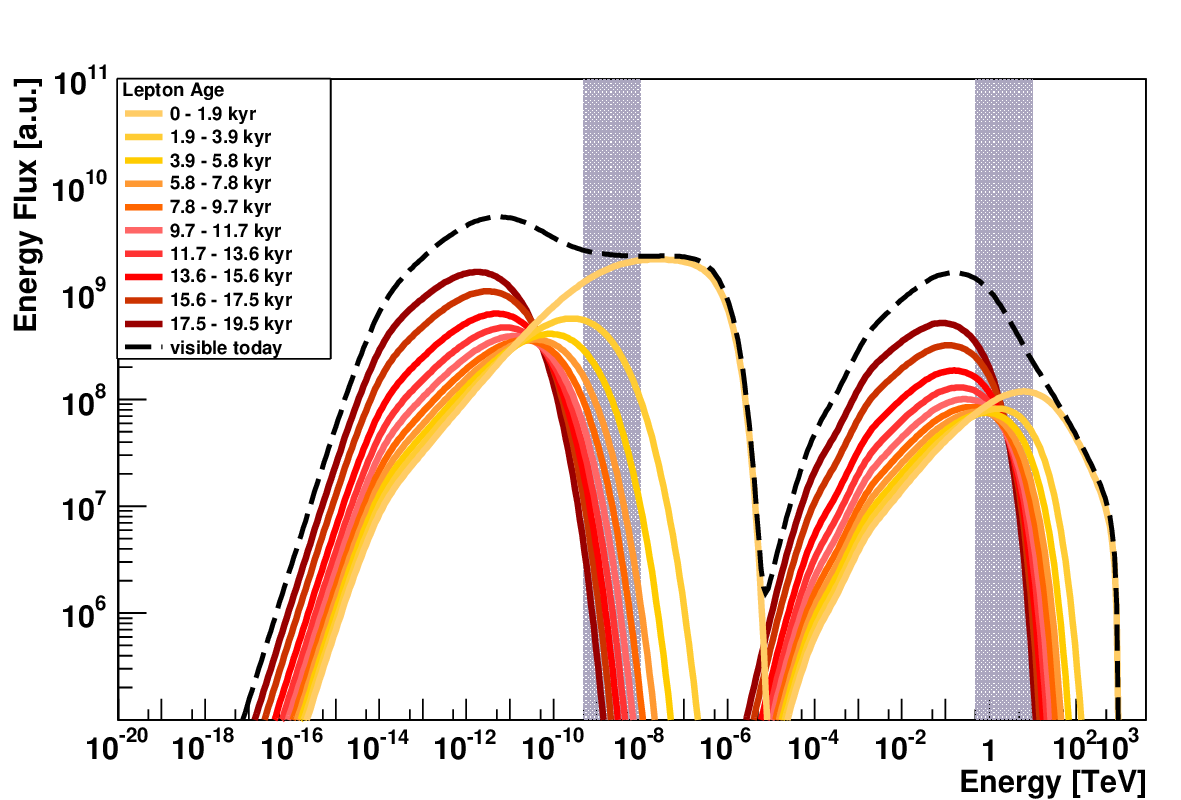}
\caption{Photon SED (black broken line) of a generic middle-aged (approximately \mbox{$20$\,kyr}) PWN decomposed into contributions by leptons from different injection epochs (solid colored lines). The same parameters as in Fig.~\ref{fig:sed_evo} were used for this example. The grey vertical bands represent the energy ranges covered by current X-ray and VHE $\gamma$-ray observatories.} 
\label{fig:contribution}
\end{figure}
Since the magnetic field strength decays strongly with time, the X-ray emission is suppressed for high PWN ages. At the same time, energy-dependent cooling effects become visible in the $\gamma$-ray band, reducing the emissions in the \vheg\ range and shifting the peak to lower energies.
\begin{figure*}[!htb]
\centering
\subfigure[MSH\,15$-$5\emph{2}]{\includegraphics[width=0.49\linewidth]{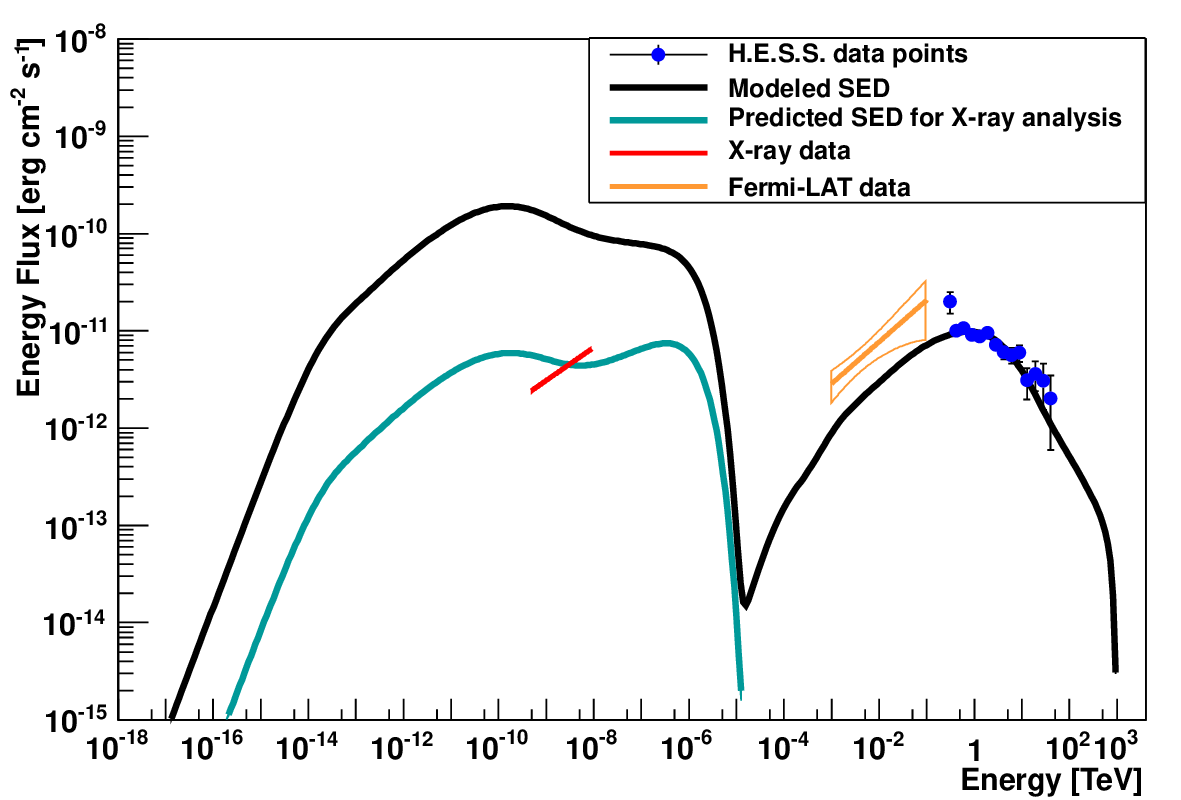} \label{fig:MSH}}
\subfigure[\hessj 1420$-$607]{\includegraphics[width=0.49\linewidth]{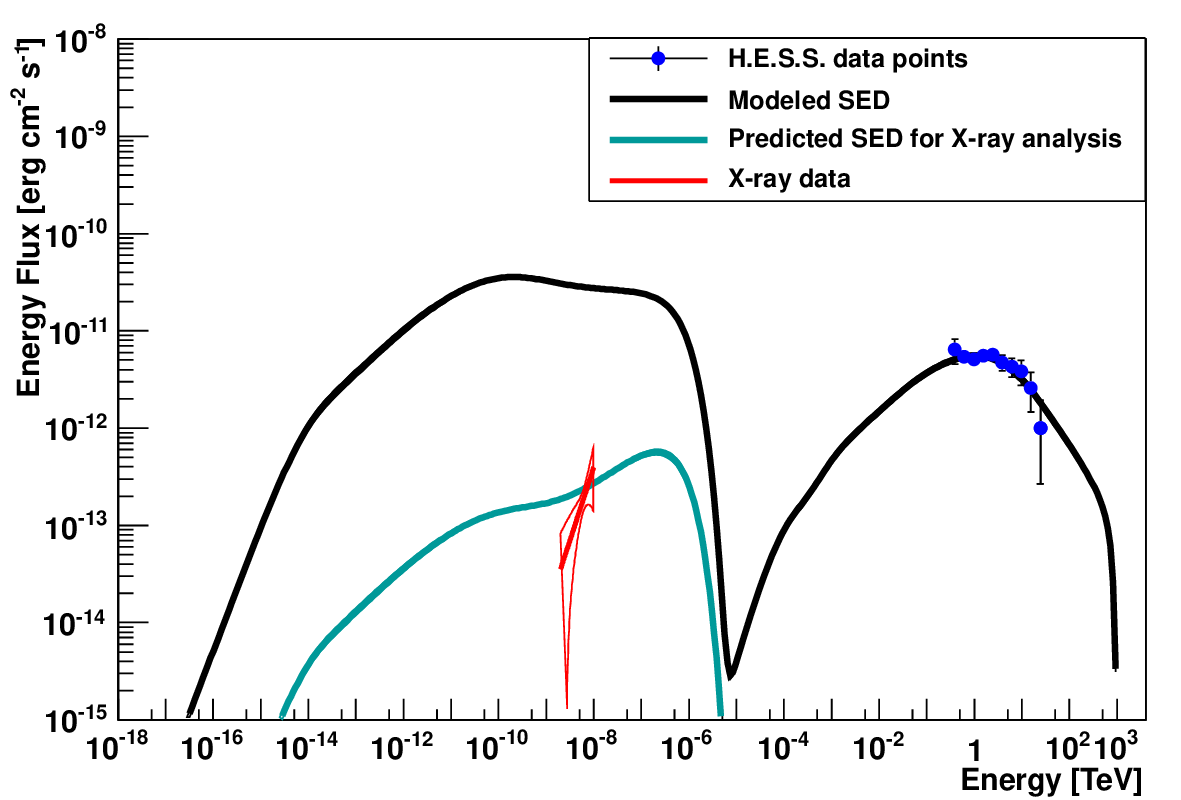} \label{fig:1420}}
\subfigure[\hessj 1825$-$137]{\includegraphics[width=0.49\linewidth]{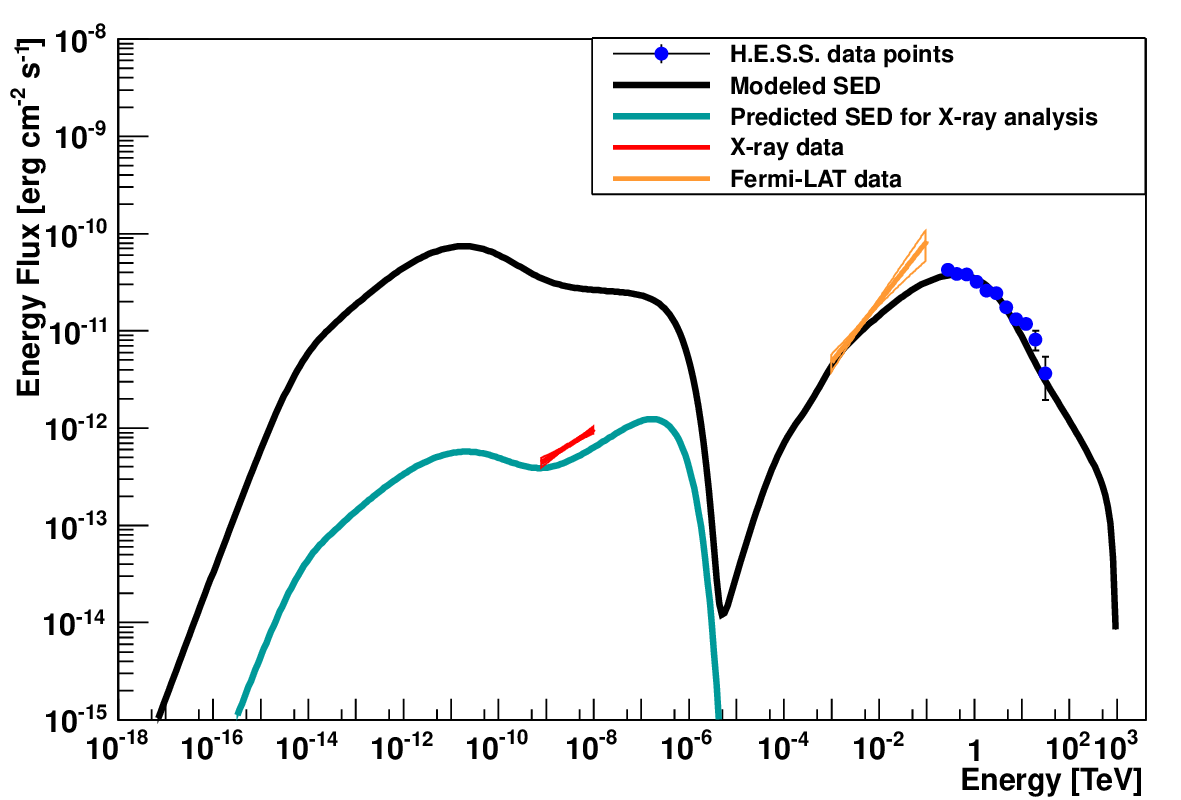} \label{fig:1825}}
\subfigure[\hessj 1837$-$069]{\includegraphics[width=0.49\linewidth]{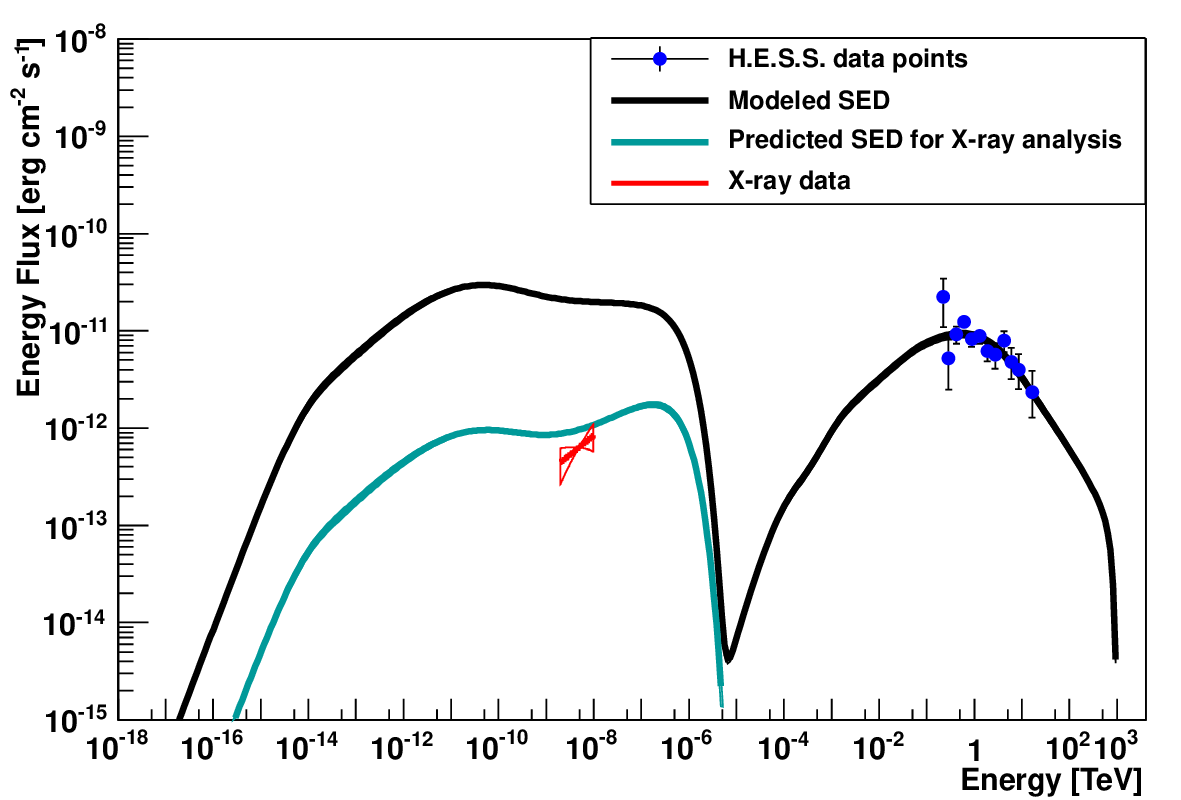} \label{fig:1837}}
\caption{Spectral energy distributions for the four sources listed in \mbox{Table \ref{tab:sources}}. The black lines show the modeled SEDs resulting from a fit to the \vheg\ data while the cyan lines denote the model prediction of the X-ray emission calculated for the published analysis regions (compare Table~\ref{tab:comp}). H.E.S.S. data (blue filled circles) are presented along with their $1\,\sigma$ statistical errors. Red lines and areas show the corresponding X-ray data with the respective error band assuming uncorrelated errors of the parameters \mbox{(see also Table \ref{tab:comp})}. If available, we included $\gamma$-ray data (orange) from the Fermi Large Area Telescope (\mbox{\emph{Fermi}-LAT}), as well. The $\gamma$-ray data are also shown along with their statistical errors. References for the \vheg\ and X-ray data can be found in \mbox{Tables \ref{tab:sources}} and \ref{tab:comp}, respectively. Fermi data for \mbox{MSH\,15$-$5\emph{2}} and for \hessj 1825$-$137 are adopted from \cite{Abdo2010} and \cite{Grondin2011}, respectively. The uncertainties on the modeled SEDs are very small and hence not visible in these plots.\label{fig:multi-sed}}
\end{figure*}
\newline 
As a second application of the model, we investigate the contribution of leptons from different injection epochs to the current photon SED, as shown in \mbox{Fig.~\ref{fig:contribution}} for our generic PWN. Summing up the SEDs from leptons of all epochs results in the emission visible today. The modeling shows that mainly the youngest, most recently injected leptons produce the X-ray emission while we observe the injection history of the pulsar in \vheg s. This can be understood as the energy of the leptons causing the synchrotron radiation in the considered X-ray regime is higher than the energy of the leptons causing the \vheg\ emission via IC scattering \citep{deJager2009}. Numerous highly energetic leptons are present in young populations. For older populations the number of highly energetic leptons has been reduced significantly by cooling processes because synchrotron and escape losses are very efficient at high energies. Thus, for the older populations the synchrotron peak is shifted to lower energies, out of the observational X-ray range.

\section{Applications}
\label{sec:res}
Based on the considerations presented above, evolved PWNe offer a potential explanation for a significant fraction of TeV sources which have remained unidentified up to now. In this scenario, an old pulsar may be surrounded by a relic TeV PWN detectable with current IACTs. However, due to the low present energy output of the pulsar the nebula does not contain enough high-energy leptons to produce a strong X-ray counterpart to the TeV PWN (compare \mbox{Figs.~\ref{fig:sed_evo} and \ref{fig:contribution}}). This view is in accordance with detailed studies presented, e.g., by \cite{deJager2009} and \cite{Mattana2009}. For some of the unidentified \vheg\ sources where an evolved PWN scenario appears likely, deep observations performed with current satellite observatories may yet reveal X-ray counterparts despite the relative weakness of the expected X-ray emission. The model presented in this work allows to select suitable candidates based on an estimate of the required exposure for a detection in the X-ray regime.
\begin{table*}[!htb]
\caption{Overview of the selected PWNe and their associated pulsars. The list is sorted by increasing characteristic age, representing different evolutionary states. The properties of the pulsars (characteristic age $\tau_\mathrm{c}$, current period $P$, current spin-down luminosity $\dot{E}$ and distance $d$) are taken from the ATNF pulsar database\footnote{\small URL: \url{http://www.atnf.csiro.au/research/pulsar/psrcat/}} \citep{Manchester2005}. References for the H.E.S.S. sources: $^{[1]}$\cite{Aharonian2005}, $^{[2]}$\cite{Aharonian2006_Kookaburra} $^{[3]}$\cite{Aharonian2006b}, $^{[4]}$\cite{Aharonian2006a}.}             
\label{tab:sources}     
\begin{center}                   
\begin{tabular}{l | c | l | c c c c}     
\hline\hline\noalign{\smallskip}           
\multirow{2}{*}{VHE Source}  & equiv. VHE source radius  & \multirow{2}{*}{Pulsar} & $\tau_\mathrm{c}$ & $P$  & $\dot{E}$  &  $d$ \\  
&  [arcmin] & &  [kyr] & [ms] & [erg $\mathrm{s}^{-1}$] & [kpc]\\
\hline                       
MSH 15$-$5{\emph 2}$^{[1]}$ & $\phantom{0}5.4$  & PSR\,B1509$-$58 & $1.55$& $151$ & $1.8\cdot 10^{37}$  & $5.81$\\
\hessj 1420$-$607$^{[2]}$  	& $\phantom{0}4.7$  	&\psrj 1420$-$6048 & $13.0$ & $68$ & $1.0\cdot 10^{37}$  & $7.65$\\ 
\hessj 1825$-$137$^{[3]}$  	& $20.7$  					&\psrj 1826$-$1334 & $21.4$& $101$ & $2.8\cdot 10^{36}$  & $4.12$\\ 
\hessj 1837$-$069$^{[4]}$  	& $\phantom{0}6.6$  	&\psrj 1838$-$0655 & $22.7$ & $70$ & $5.5\cdot 10^{36}$  & $6.6$\footnote{No distance estimate provided, we use the value proposed in \cite{Gotthelf2008} instead.}\\ 
\hline
\end{tabular}
\end{center}

\end{table*}
In order to investigate the reliability of the model we applied it to four selected PWNe, which are listed in \mbox{Table \ref{tab:sources}}. The motivation for this selection was to sample PWNe from different states of evolution for which both \vheg\ and X-ray spectra of sufficient quality are available. Since the model is radially symmetric, it was necessary to define a circular source area as an approximation to the asymmetrical Gaussian morphology fits of the published \mbox{VHE $\gamma$-ray} data. Accounting for most of the emission, the radius was chosen such that the circle covers an area equivalent to the $\sqrt{2}\,\sigma_\mathrm{ellipse}$ extent of the ellipse obtained from the VHE morphology fits ($\sigma_\mathrm{ellipse}=\sqrt{\sigma_1\sigma_2}$). The values of the used equivalent radii are included in \mbox{Table \ref{tab:sources}}. Having calculated the equivalent circular extent of the \vheg\ source, the free parameters of the model were fixed by a $\chi^2$ fit to the \vheg\ data using the \mbox{{\sc minuit}} minimization package \citep{James1975}.
Note that only statistical errors ($1\,\sigma$) of the \vheg\ data are taken into account. The optimized parameters with their uncertainties and the predictions of the \vheg\ fluxes are presented in \mbox{Tables \ref{tab:model-parameters}} and \ref{tab:vhe-results}, respectively. We used the calculated errors on the modeled parameters as well as their correlation to propagate the errors on the SEDs. Hence, it is possible to estimate the model uncertainties on the multiwavelength emission derived from the VHE $\gamma$-ray data which are in the order of $0.5\%$.
\begin{table*}[!htb]
\caption{Application of the model to the selected PWNe. The model was fit to the \vheg\ data and the fit results of the free parameters $P_0$ and $B_0$ are listed. The interval in parentheses denotes the parameter boundaries during the optimization procedure.}  
\label{tab:model-parameters}     
\begin{center}                   
\begin{tabular}{l |  c c l}     
\hline\hline\noalign{\smallskip}           
\multirow{2}{*}{Source} & $P_0$ [ms] & $B_0$ [$\mu$G] & \multirow{2}{*}{$\chi^2$/n.d.f} \\
 & $(5-P)$ & $(5-200)$ & \\
\hline                        
MSH 15$-$5{\emph 2}		& $38.9\pm 0.5$ & $104.3\pm 2.7$ & $14.5/12$\\ 
\hessj 1420$-$607 		& $33.2\pm 0.7$ & $\phantom{0}23.6\pm 1.3$ & $\phantom{0}3.5/8$\\
\hessj 1825$-$137 		& $26.0\pm 0.1$ & $\phantom{0}33.7\pm 0.4$ & $27.9/9$\\ 
\hessj 1837$-$069  		& $30.7\pm 0.8$ & $\phantom{0}24.9\pm 1.6$ & $\phantom{0}9.7/10$\\ 
\hline
\end{tabular}
\end{center}
\end{table*} 
\begin{table*}[!htb]
\caption{Comparison of predicted and measured VHE $\gamma$-ray flux for the modeled PWNe. References can be found in the caption of Table~\ref{tab:sources}.}
\label{tab:vhe-results}
\begin{center}                   
\begin{tabular}{l | c | c c }     
\hline\hline\noalign{\smallskip}           
\multirow{2}{*}{Source} & Energy threshold & \multicolumn{2}{c}{Flux above threshold\footnote{In units of 10$^{-12}$\,cm$^{-1}$\,s$^{-1}$.}}\\    
 & [TeV] & Model Prediction & Measured \\
\hline                        
MSH 15$-$5{\emph 2}	 & $0.28$ 					& $20.0$  					& $22.5$\footnote{No statistical errors provided.}   \\ 
\hessj 1420$-$607 	& $1.0\phantom{0}$   	& $\phantom{0}2.9$   & $3.0\pm 0.2$\\
\hessj 1825$-$137    	& $0.27$  					& $77.8$  					& $82.8 \pm 2.2\phantom{0}$  \\ 
\hessj 1837$-$069	 	& $0.2\phantom{0}$   	& $26.7$  					& $30.4 \pm 1.6\phantom{0}$ \\ 
\hline
\end{tabular}
\end{center}

\end{table*}
\begin{table*}[!htb]
\caption{Comparison of predicted and measured X-ray emission for the modeled PWNe. For each X-ray analysis the published energy range and the radius of the used analysis region are listed. References to the X-ray data: $^{[1]}$\cite{Schoeck2010}, $^{[2]}$\cite{Ng2005}, $^{[3]}$\cite{Uchiyama2009}, $^{[4]}$\cite{Gotthelf2008}. The uncertainties of the predicted values are in the order of $0.5\%$.}             
\label{tab:comp}     
\begin{center}                   
\begin{tabular}{l | c| c | c c | c c }     
\hline\hline\noalign{\smallskip}           
\multirow{2}{*}{Source} & \multirow{2}{*}{Analysis region} & Energy range &\multicolumn{2}{c}{$F_\mathrm{X}$\footnote{Energy fluxes are given in units of $10^{-12}\,\mathrm{erg}\,\mathrm{cm}^{-2}\,\mathrm{s}^{-1}$.}} & \multicolumn{2}{c}{Index\footnote{Photon index of a power-law model.}}\\    
 &  & [keV] & Model Prediction & Measured & Model Prediction & Measured\\
\hline                        
MSH 15$-$5{\emph 2}$^{[1]}$ 	&$30''-57''$  				& $0.5-9\phantom{0}$  	& $13.6$   	& $12.0\pm 0.3$        						& $2.1\phantom{0}$   & $1.66\pm 0.02$ \\ 
\hessj 1420$-$607$^{[2]}$ 	    &$13.5''\phantom{'}$  & $2\phantom{0.}-10$   	& $0.35$    	& $\phantom{0}0.13\pm0.03$       	& $1.77$  					& $0.5^{+1.3}_{-1.1}$ \\ 
\hessj 1825$-$137$^{[3]}$  		&$1.5'$       					& $0.8-10$ 						& $1.20$    	& $\phantom{0}1.71\pm 0.47$      	& $1.82$  					& $1.69\pm 0.09$ \\ 
\hessj 1837$-$069$^{[4]}$  		&$1.0'$       					& $2\phantom{0.}-10$   	& $1.55$    	& $\phantom{0}1.0$\footnote{No statistical errors provided.} 	& $1.87$  					& $1.6\pm 0.4$ \\ 
\hline
\end{tabular}
\end{center}

\end{table*}
The modeled SEDs are confronted with observational X-ray, $\gamma$-ray (where available) and \vheg\ data in \mbox{Fig. \ref{fig:multi-sed}}. In the \vheg\ range spectral points were available, while in X-rays and \mbox{$\gamma$-rays} the shown lines and areas correspond to the published power-law fits.\newline The comparison shows that the $\gamma$-ray and \vheg\ data are reasonably well described by the model, whereas the X-ray data are strongly overestimated. However, this is expected since the extraction region for the spectrum determination is usually much smaller in X-rays than in \vheg s. The resulting mismatch between the lepton population used for the modeling and the one observed in X-rays has to be taken into account. In the following, we assume an outflow velocity of the leptons for the innermost part of the PWN of $\nicefrac{c}{3}$ \citep[see, e.g.,][]{Kennel1984}. Starting from the given size of the X-ray spectrum extraction region, we determine the corresponding maximum age $t_\mathrm{lept.,\,max}$ of the leptons producing the emission. In the next step, we re-calculate the amount and energy distribution of the leptons contained within this region. The time scale of adiabatic energy losses for these freshly-injected leptons was adjusted accordingly ($\tau_{\mathrm{ad}} = \nicefrac{3}{2}\cdot t$, see Eq.~\ref{eq:adiab}). In order to consider projection effects in the circular region of the X-ray analysis, we added the integrated emission along the line of sight following \cite{Holler2012}. In each case we used the smallest available X-ray analysis region since the validity of the assumed outflow velocity is spatially limited to a region close to the pulsar. The resulting modified SEDs, shown in \mbox{Fig. \ref{fig:multi-sed}}, are clearly in better agreement with the observational data. A quantitative comparison between modeled and measured values of the X-ray flux is given in \mbox{Table \ref{tab:comp}}.
\section{Conclusions}
\label{sec:conclusions}
Motivated by the large number of yet unidentified \vheg\ sources suspected to be evolved PWNe, we developed a time-dependent leptonic model suitable to calculate the non-thermal emission from PWNe of different ages. The presented model allows to study the expected photon SEDs evolving with the age of the PWN. Our study yields additional support for the conception that evolved PWNe are still bright in \vheg s, while their X-ray emission is largely suppressed and hence difficult to detect.  
\newline Moreover, in this work we investigate the contribution of leptons of different epochs to the current photon SED in detail. In particular for older PWNe it is necessary to take into account the time dependence of lepton injection and cooling effects in order to explain the observed \mbox{VHE $\gamma$-ray emission}, whereas the X-ray emission (especially in the vicinity of the pulsar) is dominated by the young lepton population.
\newline 
Finally, we tested whether our model can be used to predict the X-ray flux of an unidentified source based on the \vheg\ detection. We selected four representative PWNe of different evolutionary states and fixed the free parameters of the model by a fit to observational \vheg\ data. The comparison of modeling results and observational data shows that it is possible to roughly predict the order of magnitude of the energy flux of the X-ray emission. Thus the model may facilitate the identification of evolved PWNe.

\section*{Acknowledgements}
The authors wish to acknowledge the helpful comments by members of the \hess\ collaboration. Moreover, we thank Peter Eger for the fruitful discussions concerning X-ray observations and their analysis.

\bibliography{./references}
\end{document}